\documentclass[fleqn,usenatbib]{mnras}

\usepackage{newtxtext,newtxmath}

\usepackage[T1]{fontenc}

\DeclareRobustCommand{\VAN}[3]{#2}
\let\VANthebibliography\thebibliography
\def\thebibliography{\DeclareRobustCommand{\VAN}[3]{##3}\VANthebibliography}

\usepackage{graphicx}	
\usepackage{amsmath}	
\usepackage{amssymb}	
\usepackage[normalem]{ulem}

\title[Bar and pseudobulge in Cartwheel]{Discovery of a near-infrared bar and a pseudobulge in the collisional ring galaxy Cartwheel}

\author[Barway et al. 2020]{
Sudhanshu Barway,$^{1}$\thanks{E-mail: sudhanshu.barway@iiap.res.in}
Y. D. Mayya,$^{2}$
and Aitor Robleto-Or\'us$^{3}$
\\
$^{1}$Indian Institute of Astrophysics (IIA), II Block, Koramangala, Bengaluru 560 034, India\\
$^{2}$Instituto Nacional de Astrof\'{\i}sica, \'Optica y Electr\'onica, Luis Enrique Erro  \#1, 
Tonantzintla, Puebla 72840, Mexic\\
$^{3}$Departamento de Astronom\'{\i}a, Universidad de Guanajuato, Apdo. Postal 144, Guanajuato 36000, Mexico
}

\date{Accepted XXX. Received YYY; in original form ZZZ}

\pubyear{2020}

\begin{document}
\label{firstpage}
\pagerange{\pageref{firstpage}--\pageref{lastpage}}
\maketitle

\begin{abstract}
 We report the discovery of a bar, a pseudobulge and unresolved point source in the archetype collisional ring galaxy Cartwheel using careful morphological analysis of a near-infrared (NIR) K$_s$ band image of excellent quality (seeing=0.42\arcsec) at the ESO archive. The bar is  oval-shaped with a semi-major axis length of 3.23\arcsec\
($\sim$2.09~kpc), with almost a flat light distribution along it. The bulge is almost round (ellipticity=0.21) with an effective radius of 1.62\arcsec\ ($\sim$1.05~kpc) and a Sersic index of 0.99, parameters typical of pseudobulges in late-type galaxies. The newly discovered bar is not recognisable as such in the optical images even with more than a factor of two higher spatial resolution of the Hubble Space Telescope, due to a combination of its red colour and the presence of dusty features. The observed bar and pseudobulge most likely belonged to the pre-collisional progenitor of the Cartwheel.  The discovery of a bar in an archetype collisional ring galaxy Cartwheel is the first observational evidence to confirm the prediction that bars can survive a drop-through collision along with the morphological structures like a central bulge (pseudo).
\end{abstract}



\section{Introduction}\label{sec:intro}
The Cartwheel has been the subject of multiwavelength studies due to its unusual 
morphology \citep{1941camr.book..137Z, 1977MNRAS.178..473F}, and often has been 
considered as the archetype of the class of collisional ring galaxies. These 
galaxies  are a relatively small fraction of all galaxies that have undergone 
a recent interaction \citep{2009ApJS..181..572M} and are believed to be formed 
by head-on collisions of two galaxies in which an intruder galaxy passes through 
the centre or close the center of a rotating disk of a larger galaxy, in 
the process creating an outwardly propagating density wave in the larger galaxy 
\citep{1976ApJ...209..382L, 1976ApJ...208..650T, 1996IAUS..171..337A}. As the 
expanding wave moves outwards, it triggers star formation (SF) in a circular 
ring. SF eventually stops once the wave moves forward and leaves  behind an
ageing stellar population in its wake \citep{1996ApJ...467..241H,
  1997AJ....113..201A,  2001ApJ...554..281K}. The H$\alpha$ image of
the Cartwheel suggests that most of the current SF is taking place in
29 star-forming complexes in the ring \citep{1996ApJ...467..241H},
each complex containing several compact knots at the spatial
resolution of the Hubble Space Telescope (HST).  The ring is elliptical in 
shape, with a semi-major axis length of 23~kpc and axis ratio of 0.75. The  
center of the ellipse is shifted from the nucleus by $\sim$6.5~kpc. Non-thermal radio
emission from the ageing population has been detected in narrow radial spokes
that extend to around 4~kpc inside the current ring \citep{2005ApJ...620L..35M}.

The Cartwheel has a very noticeable inner ring, which in theoretical
studies have been discussed as the second ring
\citep{1996FCPh...16..111A}. The ratio of the first ring to second
ring radii of 4.2 is in qualitative agreement with the analytical
theory of collisional ring galaxies \citep{2010MNRAS.403.1516S}. 
However, unlike the outer ring, the inner ring is gas-poor with very 
little presence of HI \citep{1996ApJ...467..241H} and H$\alpha$ \citep{1998A&A...330..881A}, 
and SF \citep{1997jena.confE.185C}. 
CO molecular line corresponding to an H$_2$ mass $\sim10^9$~M$_\odot$ 
is detected from the central parts, including the inner ring, which is found 
to be expanding at a velocity of 68.9$\pm$4.9~km\,s$^{-1}$ \citep{2015ApJ...814L...1H}. 
However, the detected CO gas in the central region is of low excitation, 
suggesting weak star-forming activity \citep{1995A&A...298..743H}.
Low SF rate of the inner ring has put some doubt on its interpretation of 
as post-collisional structure, and it could as-well be of pre-collisional origin.


\cite{1996AJ....112.1868S} speculated the presence of a faint bar inside the 
ring along with a ``D-shaped'' structure from the HST observations. However,  
their data do not conclusively confirm the bar. In a more recent paper, 
\citet{2010MNRAS.403.1516S} analysed this structure as a possible third ring. 
Numerical simulations of an intruder falling into a barred galaxy expect the 
bar to survive \citep{1997MNRAS.286..284A}. 
 
However, a comprehensive analysis of the central part in the Cartwheel 
to look for pre-collisional structures, especially a bar, is still lacking. 
The presence of a ring, and dust lanes does make it difficult to explore. 
At the same time, it should be noted that the the presence of a ring in the 
central region may possibly indicate the presence of a barlike structure as 
stellar bars are often observed encircled by a ring structure 
\citep{2005ARA&A..43..581S} and such bars are more likely to be weak in 
nature \citep{2005AJ....130..506B}. 

Observational confirmation of the bar would unambiguously prove the survival of pre-collisional structures in the central part of the Cartwheel. With this purpose, we carried out a careful structural analysis of the region inside the inner ring of the Cartwheel using Near-infrared (NIR) data. In this paper, we present the results of this study. The paper is organised as follows. Section~\ref{sec:data} describes the data. The analysis of the central region of Cartwheel and results are described in section~\ref{sec:bar}. Section~\ref{sec:discuss} is devoted to discussion and conclusions. We use a distance of 133~Mpc to the galaxy from NED\footnote{The NASA/IPAC Extragalactic Database (NED) is operated by the Jet Propulsion Laboratory, California Institute of Technology, under contract with the National Aeronautics and Space Administration.}, which corresponds to a recessional velocity of  9050~km\,s$^{-1}$ \citep{1998A&A...330..881A} for the cosmological
parameters used in NED ($\Omega_M= 0.308$, 
$\Omega_\Lambda= 0.692$ and $h_{100}=0.678$). The scale for this distance is 646~pc\,arcsec$^{-1}$.
%

\section{Data} \label{sec:data}


The Near-infrared (NIR) observations are known to reveal stellar bars
that are not seen in the optical images, the best example of such a
case is the bar of the dusty starburst M82 \citep[e.g.][]{1992ApJ...395..461T}. 
The reason for this is that NIR wavelengths probe older stellar
populations and are less affected by dust extinction and the presence 
of young stars due to recent star formation \citep{2000AJ....119..536E, 
2003AJ....125..525J, 2007ApJ...657..790M}. These facts and the speculation by 
\citet{1996AJ....112.1868S} motivated us to look for a bar in the Cartwheel using 
NIR wavelengths. For this purpose, we used the excellent quality
$K_s$-band images available in the ESO data archive\footnote{http://archive.eso.org/eso/eso\_archive\_main.html}.

The data we used belong to program ID 66.B-0666(B) in the ESO data archive,
which were taken at the VLT-U1 telescope using the ISAAC instrument in year 2000. 
An image created from this dataset has been recently used to overlay the
CO map in \citet{2015ApJ...814L...1H}, has not been used yet for a 
detailed structual analysis. The dataset included on and 
off-field (sky) images of Cartwheel, images of a standard star, in addition
to dark and flat images. The $K_s$ band images used here
were taken on three separate runs in 2000: October 15, October 18 and November 7.
Table~\ref{tab:obslog} gives the observational log with columns 1 to 5 taken
from the image headers and the last column directly measured on the image.
The dataset included 31 dithered images pointed at the object and 29 dithered 
images to sky fields on the north, south, east and west of the Cartwheel field.
Each object and sky image was formed respectively by co-adding 27 and 18 
exposures of 4 seconds each. This resulted in a total exposure time of 3348 
seconds on the Cartwheel. Airmass during all the observations remained between
1.0--1.3, and seeing between 0.35 to 0.69\arcsec, with a median seeing of 0.4\arcsec.
Each image covered a field
of view of 150\arcsec$\times$150\arcsec\ at an image scale of 0.147\arcsec/pixel.
Sky was stable to within 0.05~magnitude throughout the three runs as judged
from the variation of aperture magnitude of the in-field stars.
The data were photometrically calibrated using the observations of NIR standard star 
S294-D\footnote{https://www.eso.org/sci/observing/tools/standards/IRstandards/Photometric.html} 
\citep[K=10.596$\pm$0.004 from][]{1998AJ....116.2475P}, which were carried out during the November run. 
The standard star dataset included 5 images, each a co-add of 6 exposures of 
1.773 seconds each. Using this standard star, we get a zero point of 24.16$\pm$0.04.
This zero point is close to 24.21$\pm$0.06 obtained using 2MASS magnitudes of 2 in-frame 
stars of $K<$14.3~mag.

\begin{table}
\centering
\caption{Observational log of K-band observations}
\label{tab:obslog}
\resizebox{\columnwidth}{!}{
\begin{tabular}{llllll} 
\hline
File name & Exposure & RA & DEC & AM & Seeing \\
    (1)   &  (2)     & (3)& (4) & (5)     & (6)     \\
\hline
ISAAC.2000-10-15T06:04:09.091 & 4.0$\times$27 & 0:37:40.829 & $-$33:42:59.04 & 1.178  & 0.36 \\ 
ISAAC.2000-10-15T06:11:32.064 & 4.0$\times$27 & 0:37:41.506 & $-$33:43:00.12 & 1.198  & 0.46 \\ 
ISAAC.2000-10-15T06:14:30.998 & 4.0$\times$27 & 0:37:40.891 & $-$33:42:53.28 & 1.207  & 0.41 \\ 
ISAAC.2000-10-15T06:21:52.070 & 4.0$\times$27 & 0:37:41.100 & $-$33:43:02.64 & 1.230  & 0.41 \\ 
ISAAC.2000-10-15T06:24:53.078 & 4.0$\times$27 & 0:37:41.369 & $-$33:42:53.64 & 1.239  & 0.39 \\ 
ISAAC.2000-10-15T06:32:14.064 & 4.0$\times$27 & 0:37:40.373 & $-$33:42:54.72 & 1.265  & 0.40 \\ 
ISAAC.2000-10-15T06:35:14.035 & 4.0$\times$27 & 0:37:41.282 & $-$33:43:03.36 & 1.275  & 0.35 \\ 
ISAAC.2000-10-18T05:43:36.077 & 4.0$\times$27 & 0:37:40.939 & $-$33:42:59.76 & 1.156  & 0.40 \\ 
ISAAC.2000-10-18T05:50:59.050 & 4.0$\times$27 & 0:37:40.750 & $-$33:42:53.28 & 1.174  & 0.49 \\ 
ISAAC.2000-10-18T05:53:58.070 & 4.0$\times$27 & 0:37:41.162 & $-$33:43:00.12 & 1.182  & 0.58 \\ 
ISAAC.2000-10-18T06:01:15.082 & 4.0$\times$27 & 0:37:40.454 & $-$33:42:53.28 & 1.203  & 0.46 \\ 
ISAAC.2000-10-18T06:04:14.016 & 4.0$\times$27 & 0:37:40.373 & $-$33:43:00.48 & 1.211  & 0.37 \\ 
ISAAC.2000-10-18T06:11:32.064 & 4.0$\times$27 & 0:37:41.378 & $-$33:42:52.92 & 1.234  & 0.40 \\ 
ISAAC.2000-10-18T06:14:30.998 & 4.0$\times$27 & 0:37:41.273 & $-$33:43:03.36 & 1.244  & 0.38 \\ 
ISAAC.2000-10-18T06:27:29.981 & 4.0$\times$27 & 0:37:40.260 & $-$33:42:57.96 & 1.291  & 0.42 \\ 
ISAAC.2000-10-18T06:37:48.086 & 4.0$\times$27 & 0:37:41.244 & $-$33:42:57.96 & 1.332  & 0.69 \\ 
ISAAC.2000-11-07T02:27:30.038 & 4.0$\times$27 & 0:37:40.930 & $-$33:43:00.84 & 1.014  & 0.40 \\
ISAAC.2000-11-07T02:34:52.061 & 4.0$\times$27 & 0:37:40.399 & $-$33:42:56.16 & 1.016  & 0.41 \\
ISAAC.2000-11-07T02:37:50.995 & 4.0$\times$27 & 0:37:41.496 & $-$33:42:58.68 & 1.017  & 0.41 \\
ISAAC.2000-11-07T02:45:11.030 & 4.0$\times$27 & 0:37:41.544 & $-$33:43:06.96 & 1.020  & 0.44 \\
ISAAC.2000-11-07T02:48:10.051 & 4.0$\times$27 & 0:37:41.246 & $-$33:43:16.68 & 1.022  & 0.49 \\
ISAAC.2000-11-07T02:55:27.062 & 4.0$\times$27 & 0:37:41.549 & $-$33:43:21.00 & 1.026  & 0.55 \\
ISAAC.2000-11-07T02:58:26.083 & 4.0$\times$27 & 0:37:42.036 & $-$33:43:18.48 & 1.028  & 0.49 \\
ISAAC.2000-11-07T03:06:55.066 & 4.0$\times$27 & 0:37:40.939 & $-$33:42:59.76 & 1.035  & 0.40 \\
ISAAC.2000-11-07T03:14:15.101 & 4.0$\times$27 & 0:37:40.375 & $-$33:42:59.76 & 1.041  & 0.36 \\
ISAAC.2000-11-07T03:17:14.035 & 4.0$\times$27 & 0:37:41.153 & $-$33:43:04.08 & 1.044  & 0.35 \\
ISAAC.2000-11-07T03:24:37.008 & 4.0$\times$27 & 0:37:40.666 & $-$33:43:06.24 & 1.052  & 0.37 \\
ISAAC.2000-11-07T03:27:35.078 & 4.0$\times$27 & 0:37:41.230 & $-$33:42:54.00 & 1.055  & 0.36 \\
ISAAC.2000-11-07T03:34:57.965 & 4.0$\times$27 & 0:37:41.462 & $-$33:43:01.92 & 1.064  & 0.42 \\
ISAAC.2000-11-07T03:37:57.072 & 4.0$\times$27 & 0:37:40.320 & $-$33:42:54.72 & 1.068  & 0.38 \\
ISAAC.2000-11-07T03:45:18.058 & 4.0$\times$27 & 0:37:41.398 & $-$33:43:06.96 & 1.079  & 0.37 \\
\hline
\end{tabular} 
}
Notes on columns: (1) File name in the ESO data archive formed as a combination of
instrument (ISAAC) followed the starting universal time in standard fits keyword format.
(2) Each exposure consists of co-addition of 27 frames of 4 second each. (3--4) The equitorial 
coordinates of the telescope pointing. (5) AIRMASS at the start of each observation.
(6) Seeing measured using a star in each frame.
\end{table}
%

\begin{figure*}
\rotatebox{0}{\includegraphics[width=16cm]{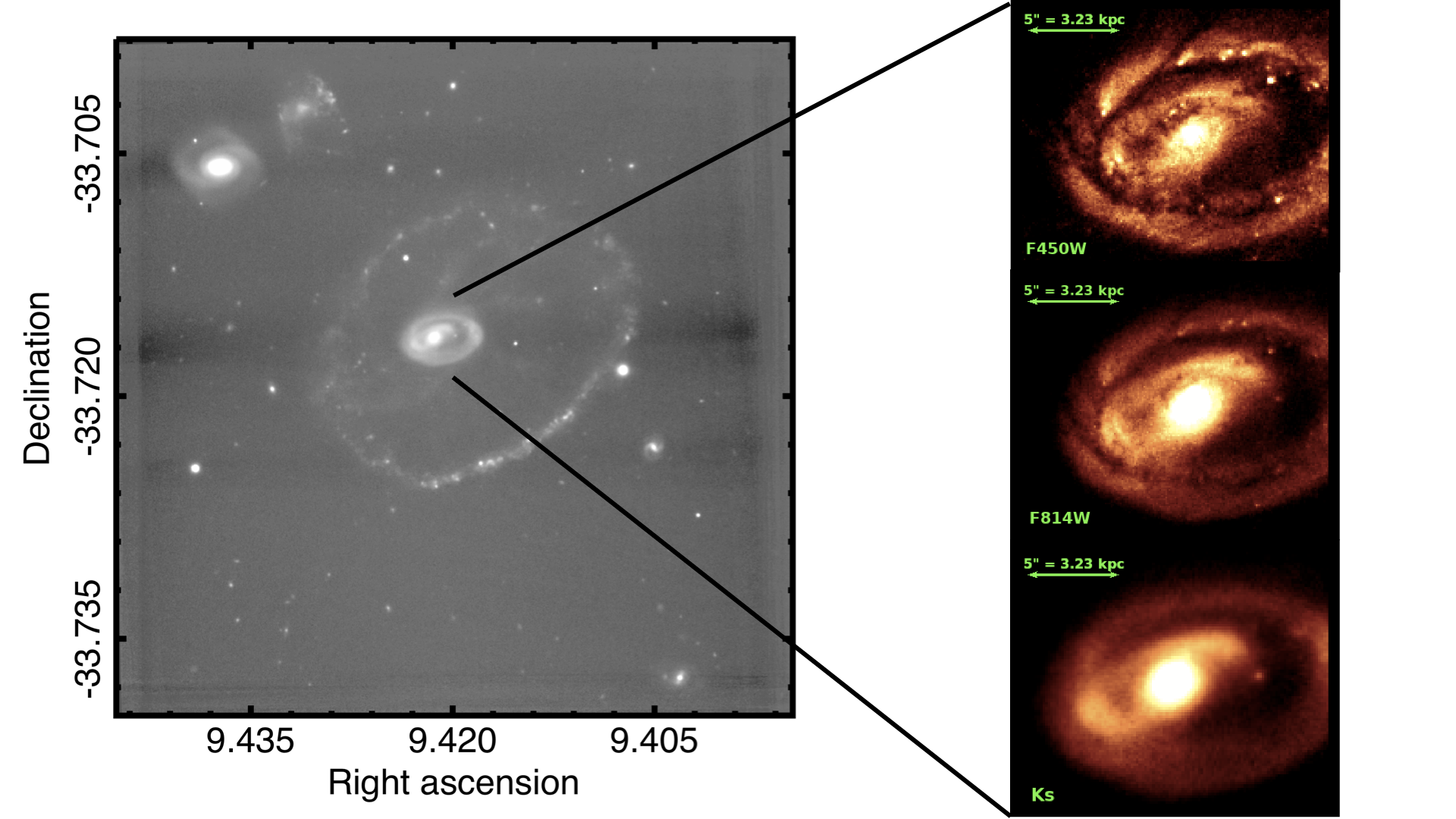}}
\caption{Cartwheel galaxy image in the K$_s$ band, where north is up and east is to the left. 
The central part is shown enlarged to the right to provide visual illustration in HST/WFPC2 F450W (top), F814W(middle) and  K$_s$ band (bottom) for the presence of a bar and a central bulge, all inside the inner ring. }
\label{fig:f1} 
\end{figure*}


We followed the standard NIR image reduction techniques, that included 
preparing master dark images and flat images. Exposure time matched dark images 
are subtracted from individual flat-field, Cartwheel and sky images.
Dark-subtracted flat-field images for each run are combined using median
algorithm, which were normalized by the median count of the combined image.
Individual dark-subtracted Cartwheel and sky frames are divided by the
normalized flat-field. The resulting sky frames are then combined using median
algorithm, which is then subtracted from each Cartwheel image.
Each of the resulting 31 Cartwheel images were visually 
checked to be of good quality by displaying them on a ds9 monitor.
We selected 10 non-saturated stars in the field, which we used to measure
the seeing and photometric quality of each individual image, in addition 
to getting relative shifts to facilitate registering the dithered images.
Astrometrically registered images were combined using the median algorithm.
We measured a Full Width at Half Maximum (FWHM) of 0.43\arcsec\ for the
Point Spread Function (PSF) in the combined image.
We also generated a combined image of 7 images with the best seeing, which
resulted in an image of FWHM=0.39\arcsec. We recall that the images with the
highest spatial resolution available for the Cartwheel are from the HST/WFPC2 
camera in the F450W and F814W-band which have a PSF of 0.17\arcsec\ FWHM, hence
$\sim$2.3--2.6 times better than our images.

The entire data reduction was carried out using the tasks in 
{\sc IRAF}\footnote{IRAF is distributed by the National Optical Astronomy
Observatories, which are operated by the Association of Universities for
Research in Astronomy, Inc., under cooperative agreement with the National
Science Foundation.}, as described in \citet{2005AJ....129..630B}. 
The images were astrometrised using the coordinates of 2MASS stars in the 
image with the help of tasks {\it ccmap} and {\it wregister}.
The final image is displayed in Figure~\ref{fig:f1}.



\section{Identifying bar in Cartwheel} \label{sec:bar}

Bars can be identified using various criteria listed in 
literature \citep{2000AJ....119..536E, 2002MNRAS.336.1281W,
2007ApJ...657..790M, 2005MNRAS.362.1319L, 2009ApJ...696..411W}. 
Morphological classification of galaxies, including the inference of 
a bar, has been carried out historically by visual inspection of 
images \citep{1963ApJS....8...31D}.  Even in the current digital 
era, classification by eye continues to be the most powerful 
tool \citep{2013pss6.book....1B}. Using the SDSS imaging survey, 
a visual morphological classification of galaxies is carried out 
as a part of the Galaxy Zoo projects \citep{2008MNRAS.389.1179L, 2013MNRAS.435.2835W} 
in which citizen scientists were asked to give detailed information about 
the visual appearance of galaxies along with features such as bars in 
the galaxies. These bar classifications were used in several  Galaxy 
Zoo studies of barred galaxies to investigate the role of bars to understand 
the formation and evolution scenarios of galaxies 
\citep{2012MNRAS.424.2180M, 2012MNRAS.423.1485S, 2018MNRAS.473.4731K}. 

 We show the K$_s$ band image of the Cartwheel in 
the left panel of Figure \ref{fig:f1}. The right panel shows a close-ups of 
the central part showing only the inner ring in HST/WFPC2 F450W (top panel) 
and  F814W (middle panel), in addition to that in the K$_s$ band (bottom panel).
The images have been displayed with suitable intensity scaling so as to allow
a direct visual apreciation of the prominant features in the central region. 
A diagonal linear structure (bar) 
and a central bright component, which is a candidate either for an unresolved
nucleus or a bulge, are clearly noticeable in the K$_s$ band image. On the other
hand, in the optical bands the bar is not apparent, in spite of more than a 
factor of two higher spatial resolution offered by the HST. Instead, the most prominant
feature is a spiral-like structure that seems to emerge from the south-east
end of the bar, almost at 90$^\circ$ to it. This spiral feature continues north-ward
but bends again sharply towards the north-west end of the bar, merging with a
short spiral feature emerging from the opposite end of the bar.
This overall morphology is the ``D-shaped'' structure pointed out by \citet{1996AJ....112.1868S}.
These authors had also noted the presence of a network of dust lanes including 
large cometary structures in the inner ring, which all can be seen in
the optical images, especially in the F450W band. The spiral is seen in
the K$_s$-band image, but at a lesser contrast than the bar, whereas
all the dust-related structures are absent in the near infrared image.

\begin{figure*}
\rotatebox{0}{\includegraphics[scale= 0.29]{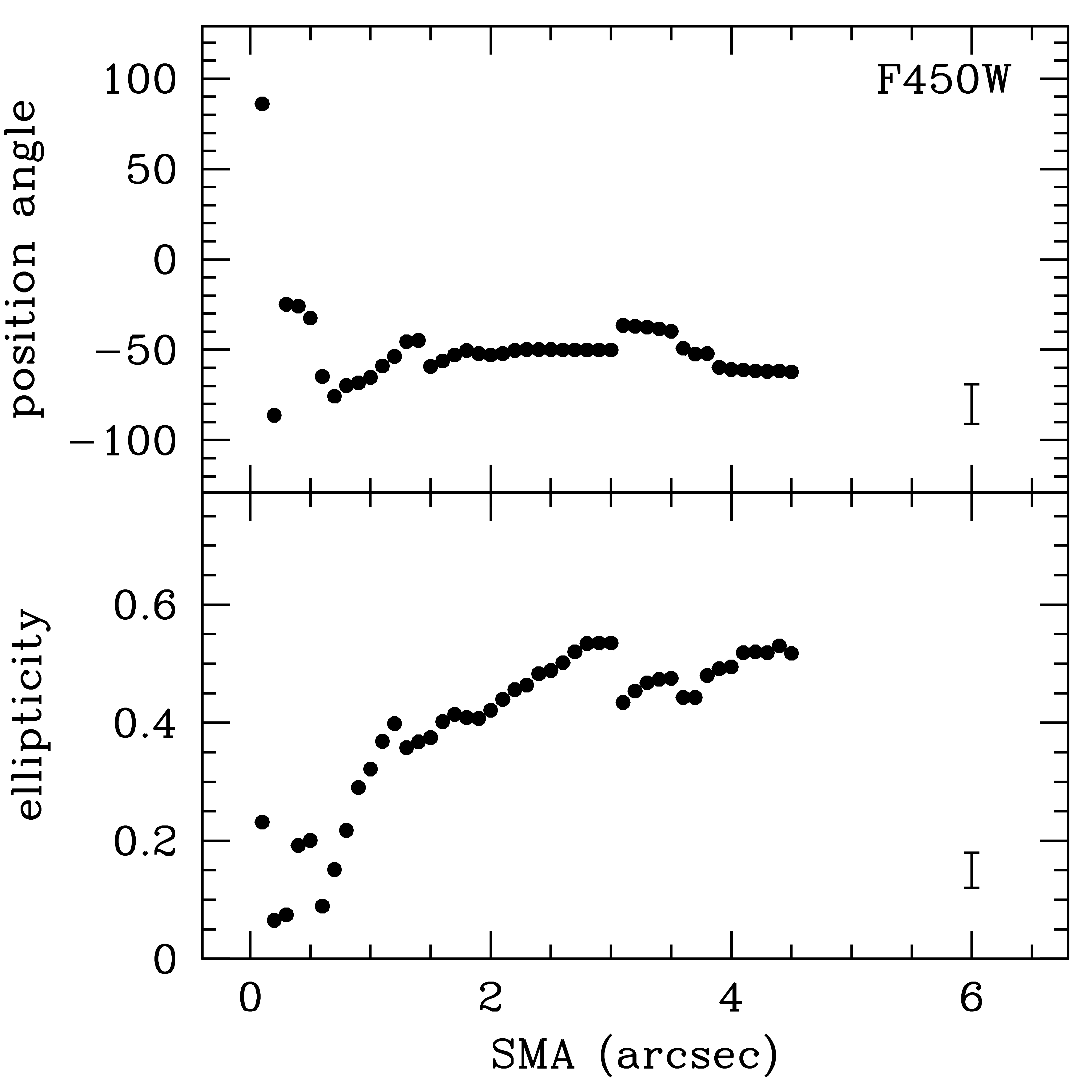}}%
\rotatebox{0}{\includegraphics[scale= 0.29]{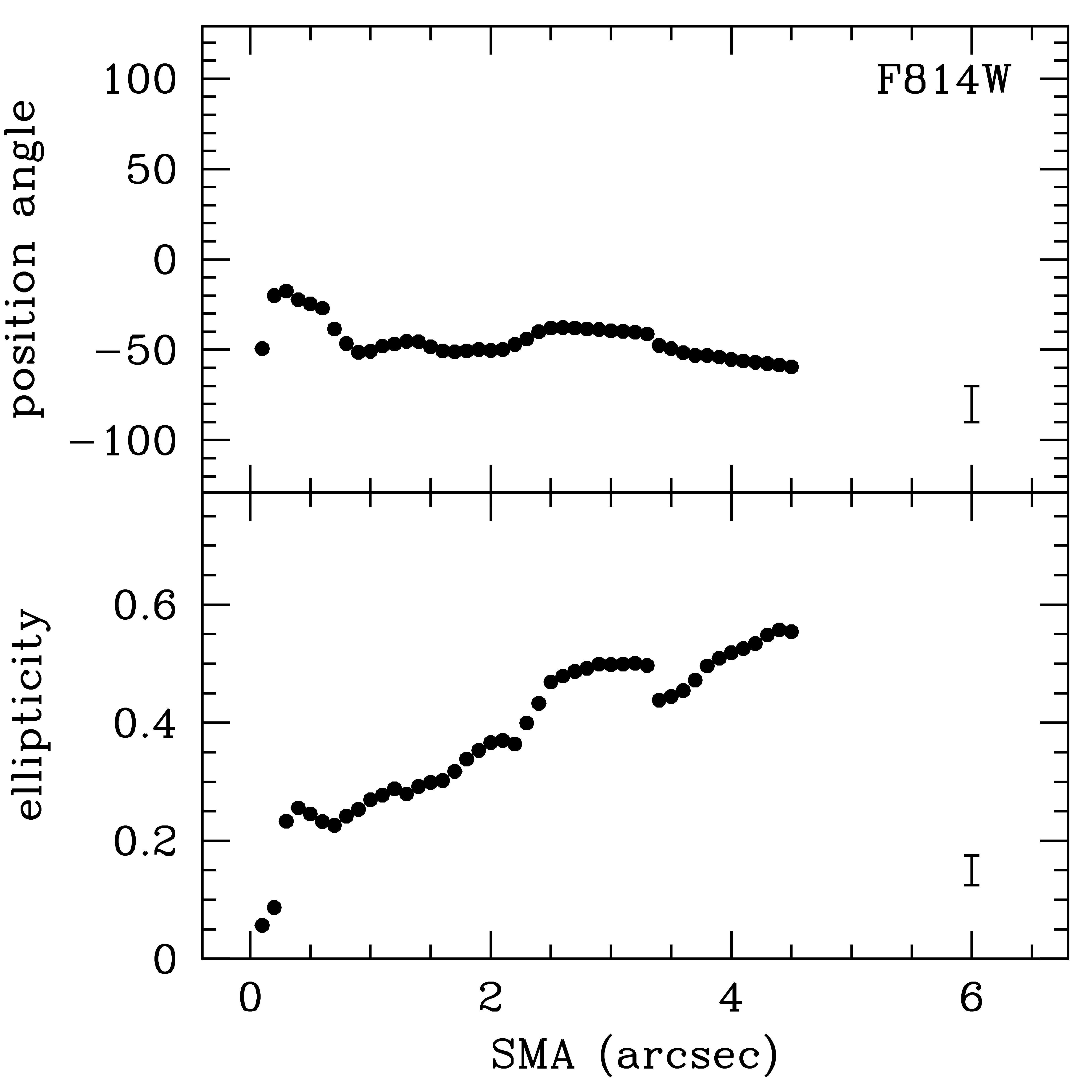}}%
\rotatebox{0}{\includegraphics[scale= 0.29]{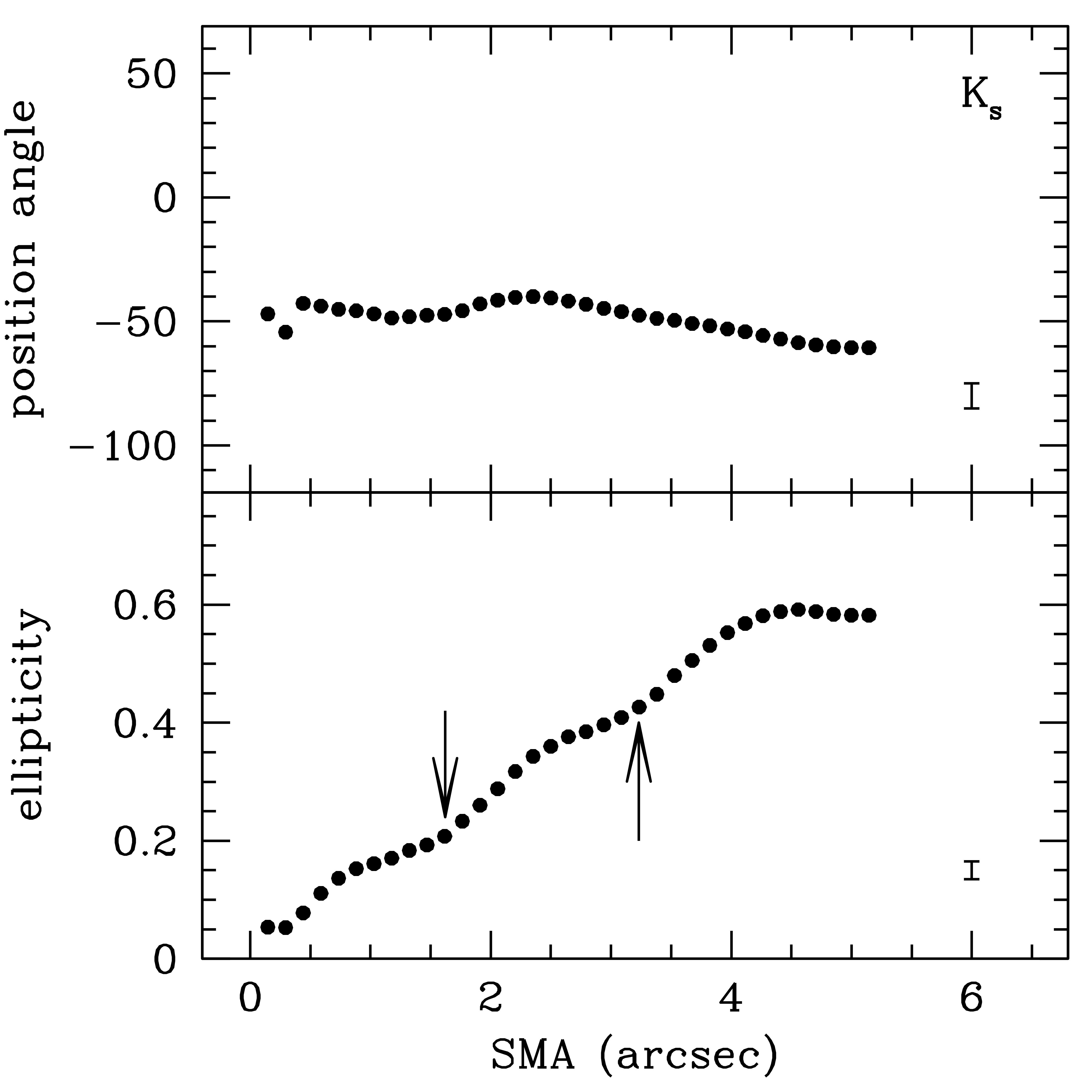}}
\caption{Results of ellipse fitting:  position angle (top) and ellipticity (bottom)  profiles of the best-fitted ellipse plotted as a function of the semi-major axis of the ellipse for HST-WFPC2 F450W (left), HST-WFPC2 F814W (middle) and K$_s$ band (right). In the  K$_s$ band ellipticity profile,  the position of the end of the bar is shown by an upward pointing arrow and the position of the end of the bulge is shown by a downward pointing arrow. The typical error bar for position angle and ellipticity profile is shown on bottom right corner of the plot.}
\label{fig:f2} 
\end{figure*}



\begin{figure*}
\rotatebox{0}{\includegraphics[width=17cm]{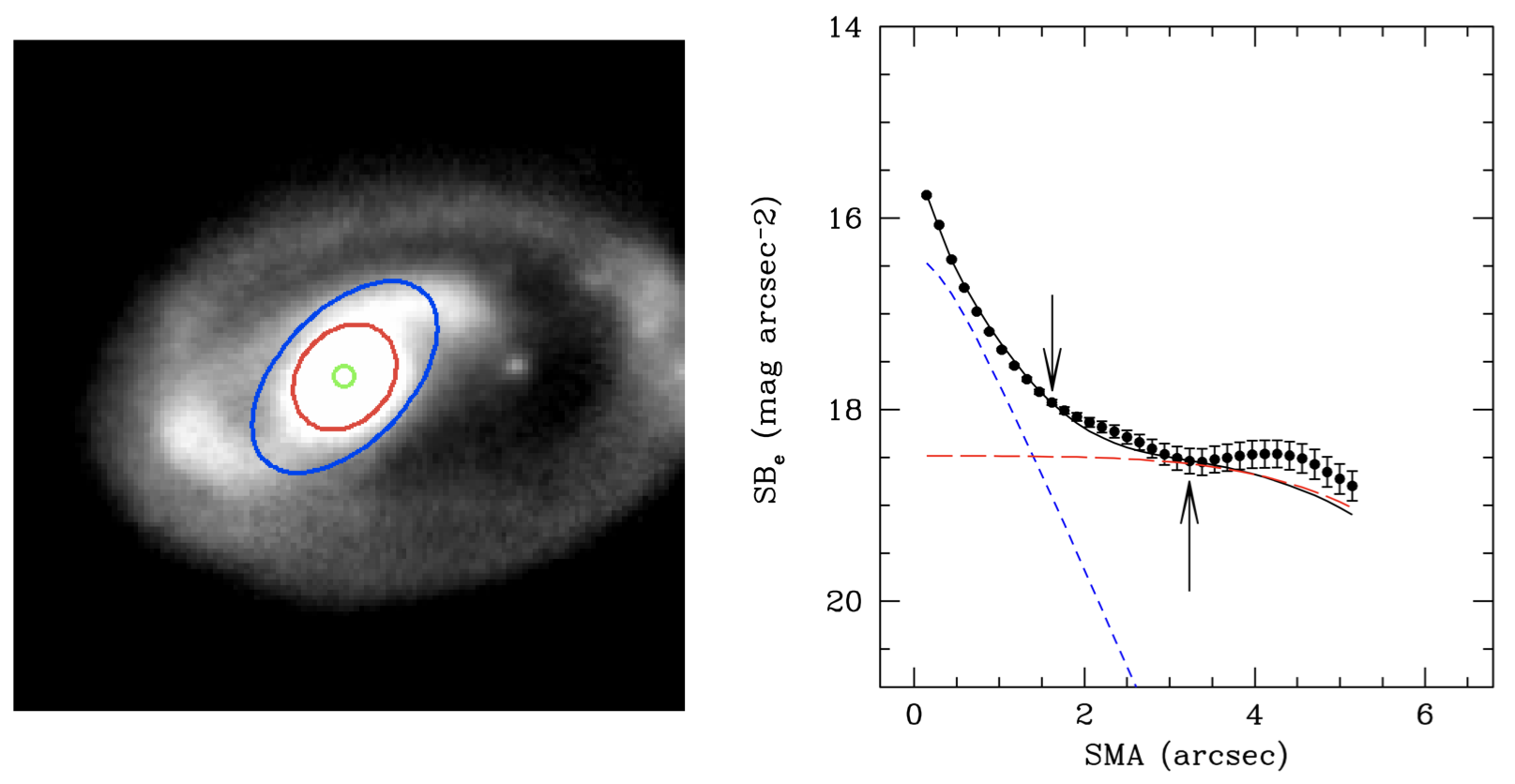}}
\caption{{\bf Left :}  Figure identifying the morphological components in the inner part of the Cartwheel.  A $K_s$-band image showing the bulge (shown by a red ellipse), bar (shown by a blue ellipse), and unresolved point source (shown by a red circle). The parameters of the displayed ellipses/circle are those tabulated in Table~\ref{tab:bartab} for {\it ellipse} fitting.  {\bf Right :} Results of ellipse fitting. Azimuthally averaged surface brightness of the best-fitted ellipse plotted as a function of the semi-major axis of the ellipse (solid dots). The surface brightness profile is decomposed into a point source and two Sersic components, an inner bulge (blue dashed) and an outer bar (red dashed). The sum of these three components is shown by a black solid line. The position of the end of the bar is shown by an upward pointing arrow and the position of the end of the bulge is shown by a downward pointing arrow.}
\label{fig:f3} 
\end{figure*}

\subsection{Ellipse fitting} \label{sec:efit}
The most widely used quantitative technique to identify a bar is the ellipse-fitting 
method. The bars are features where the position angle (PA) of the isophotes remains 
constant and the ellipticity ($\epsilon=1-b/a$; $a$ and $b$ are the semi-major and 
minor axes, respectively)  increases steadily to a global maximum 
\citep{1995A&AS..111..115W, 2007ApJ...659.1176M, 2008ApJ...675.1194B, 2009A&A...495..491A}. 
Also, the ellipticity of the bar is directly a measure of the strength of the 
bar \citep{2005MNRAS.364..283E}. This criterion is based on the work done by 
\citet{1992MNRAS.259..328A} who showed that in the bar region,  the bar-supporting 
orbits (also known as the  {\it x1} family of orbits) can be modelled by concentric 
ellipses with a constant position angle as a function of the semi-major axis.  Sometimes 
the second criterion also used in which the ellipticity drops by 0.1 coupled with the 
change in position angle at the region of the transition from bar to disc as at this point 
there is a transition from highly eccentric {\it x1} orbits at the bar end to the more 
circular orbits near the disc \citep{2007ApJ...659.1176M, 2004ApJ...615L.105J}. For our 
work, we use the former criterion as in the inner region of Cartwheel, it is non-trivial 
to identify the transition from bar to disc region due to the presence of the inner ring and
the spiral arm.

We used the ellipse-fitting task ELLIPSE in the STSDAS package available in IRAF to 
fit the part interior to the inner ring of HST-WFPC2 optical and  K$_s$-band images. Following 
\cite{2005AJ....129..630B}, we started the fitting few arcseconds from the centre of the 
Cartwheel and stopped at the isophotes where the rings start appearing. We have identified 
the stars, bad pixels and other hidden features etc. during the first round of fitting and 
masked in the subsequent run.   All the parameters, including the centre of the ellipse, were 
allowed to vary during the fitting. As the majority of the inner region of the Cartwheel contains 
complex dust features, accurate estimation of the initial values of  ELLIPSE task parameters 
was not possible. To minimize the error in this estimate, we first carried out isophote fitting 
for  K$_s$  band image which is least affected by dust compared with optical images. The parameters 
thus determined were used in the second run and the above described ellipse-fitting procedure 
was repeated. The same parameters were used to fit ellipses to the isophotes in optical bands. 
We repeated the fitting process with different starting major-axis lengths and different forms 
of sampling to check the stability of the extracted parameters.  This fitting procedure gives 
the ellipticity, position angle and the mean intensity (surface brightness) along the major axis 
for the best-fit ellipse, as a function of the semimajor-axis length.

 In Figure \ref{fig:f2}, we show the position angle (top panel) and ellipticity (bottom panel)  
for HST WPC2 F450W (left), F814W (middle) and K$_s$ (right) band images.  The position angle and 
ellipticity profiles for F450W and F814W is rather disturbed with several breaks in the profiles. 
This is due to the presence of dust in the inner region of Cartwheel which causes 
the departure of the isophotes from the best-fit ellipse.  Therefore,  the 
position angle and ellipticity profiles from F450W and F814W band images do not help 
to identify bar. However, the dust hardly affects the morphology of the K$_s$-band 
image giving us a smooth position angle and ellipticity profiles that we use to characterise the bar.

The isophotes are nearly circular in the centre upto $\sim$0.29\arcsec, 
which indicates the presence of an unresolved central component --- a nucleus.
Beyond this point ellipticity gradually increases with very little variation 
in the position angle. Both the ellipticity and the position angle show a 
simultaneous change at $\sim$1.62\arcsec (1.05 kpc; shown by the downward pointing arrow),
indicating the presence of a central bulge-like component. The position angle continues to vary, albeit rather little, all along the plotted 
semi-major axis length of 5\arcsec. The gradually increasing ellipticity shows a tendency to 
flatten at $\sim$3.2\arcsec (2.09 kpc; shown by the upward pointing arrow). Such a 
behaviour is the characteristic signature of a bar \citep{2007ApJ...657..790M}.  
In some cases, prominent spiral arms starting at the two ends of the bar can cause 
ellipticitity to keep rising beyond the end of the bar. These spiral arms also increase the radial region of a nearly constant position 
angle as seen in the position angle profile. Visual examination of the image 
suggests a spiral-like structure emerging from both sides of the bar at this radius. 
Thus, the bar ends at $\sim$3.2\arcsec\ (2.09 kpc) and the ellipticity increase 
beyond this point is due to the contamination from the spiral arm. This strategy 
used by us to measure the {\it length of the bar} is consistent with the 
often-adopted method of measuring the bar length in normal galaxies as the point 
where the ellipticity profile reaches a maximum  \citep{2005MNRAS.364..283E, 2007MNRAS.381..943G, 2007MNRAS.381..401L}.
In the left panel of Figure~\ref{fig:f3}, we superpose the ellipses corresponding
to the nucleus, bulge and the bar on the $K_s$-band image. The ellipses corresponding to the 
bulge and bar enclose very well the visually identifiable morphology of these components.

Another important bar characteristic apart from "bar length" is 
a `strength' of a bar. There are various methods adopted in the literature to 
determine the bar strength such as maximum ellipticity of the isophotal 
ellipticity profile \citep{2007ApJ...659.1176M,  2009A&A...495..491A, 
2015A&A...582A..86H}, using bar torques \citep{2001ApJ...550..243B}, the maximum 
amplitude of the m = 2 Fourier mode \citep{2002MNRAS.330...35A, 2005MNRAS.362.1319L} 
or two-dimensional fast Fourier transform (FFT) method \citep{2017A&A...601A.132G}.  
\citet{2000AJ....120.2835A} introduced the parameter $f_{bar}$ to derive the bar 
strength, that uses the axial ratio $(b/a)$ of the bar. The parameter $f_{bar}$ 
has a value between zero (unbarred) to close to unity (strong bar) and can be 
used to identify weak and strong bars as a function of galaxy morphology. We 
have adopted the modified form of  parameter $f_{bar}$ as given in 
\citet{2009A&A...495..491A} 
%
\begin{equation} 
f_{\rm bar} = \frac{2}{\pi} \left(\arctan (1-\epsilon_{\rm bar})^{-1/2}
  - \arctan (1-\epsilon_{\rm bar})^{+1/2} \right), 
\end{equation}
%
where the ellipticity ($\epsilon_{\rm bar}$) is derived from the  maximum value of the ellipticity profile along the bar,
which usually happens at the end of the bar. We determined $\epsilon_{\rm bar}$=0.43  
for the Cartwheel bar which gives the $f_{\rm bar}$=0.36, a value typically 
found for late-type spiral galaxies with strong bar \citep{2009A&A...495..491A}. 
The bar length and strength that we find for Cartwheel is also similar for the 
late-type galaxies reported by \citet{2016A&A...596A..84D} using the 
3.6\,$\micron$ imaging from the Spitzer Survey of Stellar Structure in Galaxies ($S^4G$). 

We would like to point out that measuring the length or ellipticity of a 
bar in galaxies is a non-trivial task as the surface brightness profile gets 
affected at the end of the bar by several known morphological sub-structures 
such as dust lane, spiral arm, ring etc. \citep{2005MNRAS.364..283E} along with 
deprojection effects. For a careful characterization of a bar, it is important 
to deproject the galaxy image to face on to have reliable measurements of length 
and ellipticity of the bar. However, deprojection techniques systematically 
introduce uncertainty in the measurements of bar length and ellipticity if the 
parameters used are derived incorrectly \citep{2007MNRAS.381..943G, 2014ApJ...791...11Z}. 
To deproject a galaxy image the techniques employed in the literature use the 
following parameters --- position angle of the 25 mag arcsec$^{-2}$ isophote 
in the B-band and the inclination angle of the disc of the galaxy ({\it i}) 
to the plane of the sky. In many galaxies, these parameters are difficult to 
measure and the Cartwheel is no exception due to its collisional origin and 
the presence of sub-structures mentioned above. We, therefore, do not apply 
depoejection for measuring the length and ellipticity of the Cartwheel bar. 


\begin{table}
\centering
\caption{Bar \& bulge parameters for Cartwheel.}
\label{tab:bartab}
\resizebox{\columnwidth}{!}{\begin{tabular}{llcrrcrrc} 
\hline
  & \multicolumn{3}{c}{ELLIPSE} & \multicolumn{4}{c}{GALFIT} \\
\hline
 & \multicolumn{1}{c}{$a$} & \multicolumn{1}{c}{$\epsilon=$} & \multicolumn{1}{c}{PA} &
\multicolumn{1}{c}{r$_e$} & \multicolumn{1}{c}{n} &
                        \multicolumn{1}{c}{b/a}
                          &  \multicolumn{1}{c}{PA}  & \multicolumn{1}{c}{$log (M_{*} / M_\odot)$} \\
\hline
 & \multicolumn{1}{c}{kpc} &  \multicolumn{1}{c}{1-b/a} & \multicolumn{1}{c}{$^\circ$} &
\multicolumn{1}{c}{kpc} & &  &  \multicolumn{1}{c}{$^\circ$} & \\
\hline
Bar     & 2.09 & 0.43 & $-$48 & 2.61 & 0.23 & 0.60 & $-$57 & 9.70 \\
Bulge   & 1.05 & 0.21 & $-$47 & 0.56 & 0.99 & 0.72 & $-$45 & 9.17 \\
\hline
\end{tabular}}
\end{table}

\subsection{Two-dimensional image decomposition} \label{sec:2dfit}

In order to obtain a completely independent measure of the central 
component, and the bar, we adopted the technique of two-dimensional 
decomposition of galaxy light in our K$_s$ band image using the GALFIT 
\citep{2002AJ....124..266P} 2D decomposition code.  The GALFIT uses the 
fast Marquardt-Levenberg algorithm to minimise the $\chi^2$ between the
observed image and the PSF-convolved sum of a variety of model components. 
Parameters of the model with the minimum $\chi^2$ are considered
optimal values. A GALFIT run requires the following inputs: an input image, a mask image, 
a PSF image and a configuration file containing a description of the model 
to be fitted. The PSF was obtained by identifying foreground stars in the 
K$_s$ band image and by fitting Gaussian using the PSF task available in the 
DAOPHOT package within IRAF. The mask image is created using the IRAF ELLIPSE 
task. We modelled both the bulge and the bar using the Sersic function 
\citep{1963BAAA....6...41S}. The central point source was modelled as an 
unresolved point source convolved with the PSF used for GALFIT fitting, 
thus there is no prescribed analytical functional 
form and the only parameter fitted by the GALFIT is total magnitude. 

The resulting surface brightness profiles are shown in the right panel of Figure \ref{fig:f3}. 
The parameters of these fits are compared to that obtained by the ellipse fitting 
in Table~\ref{tab:bartab}. The central component fitted as unresolved point source 
has a total magnitude of 17.66.  The bulge component has a  Sersic index of $n=0.99$, 
and a effective radius of 0.56~kpc (0.89\arcsec), whereas the second Sersic component has 
a much flatter light distribution ($n=0.23$) and is longer, and represents the bar. 
Sersic index of $n=1.0$ corresponds to an exponential profile, and hence the bulge
in the Cartwheel is an exponential bulge. The axis ratio and the PA of the two components are in agreement 
with those derived for these two components using the ellipse fits. \cite{2011MNRAS.415.3308G} 
has found that the bar length (obtained from isophotal semi-major axis length) is well 
correlated  with bar effective radius $r_{\rm e}$. However, in the Cartwheel, 
$r_{\rm e}=2.61$~kpc obtained from GALFIT is larger than the bar length (2.09~kpc) 
obtained by ellipse fitting, which was expected given the interference from the spiral 
structure, which was not modelled in the GALFIT. We will use the semi-major axis measured 
from ellipse fitting as the bar length, henceforth. The $b/a$ parameter  from Table~\ref{tab:bartab}  
indicates that the bar is somewhat oval shaped. The Sersic index much lower than 1 suggests that 
the intensity varies very little along the bar. Also, the results given in Table~\ref{tab:bartab} for the two Sersic functions we used, 
corresponding to bar and bulge, indicate that the bulge is less massive than the bar.
The mass of the individual components is obtained using M/L of 0.6 in $K_s$ band 
\citep{2014AJ....148...77M}. Note that we have not applied  reddening corrections due to 
Galactic extinction as it is negligible in the K$_s$ band. 

The errors estimated on the recovered parameters by GALFIT takes into account
only the poisson noise due to the source brightness, which is small as compared
to real errors due to possible PSF variation in the image, correlated noise due 
to parameter degeneracy, and most importantly the errors in estimating the sky 
background which is one of the main sources of errors  \citep{2010AJ....139.2097P}. 
Apart from these, the unmodelled spiral arms cause additional uncertainties on
the recovered parameters. As it is non-trivial to estimate the errors 
due to all these factors, in Table~\ref{tab:bartab}, we present values 
without their corresponding errors. Notwithstanding, GALFIT has been
useful to confirm the presence of a bar and bulge that we find from ellipse 
fitting analysis. For a detailed discussion on error estimation using GALFIT, we recommend the 
reader to \citet{2018MNRAS.473.4731K} and \citet{2013MNRAS.435..623V}.


\section{Discussion and conclusions} \label{sec:discuss}


In spite of a lot of interest in the Cartwheel, the morphology of its host 
galaxy is still under deliberation. From our analysis of NIR K$_s$ band images 
presented in the previous sections, we conclude that that Cartwheel has an 
oval-shaped bar, a pseudobulge and an unresolved nucleus.

The derived bar length and bar strength suggest that the discovered bar
is a strong bar. The fundamental question that needs to answered is whether 
these structures formed 
after the collision or were they already there at the time of the collision?
\citet{2008AN....329..948M} carried out N-body simulations of collisional
ring galaxies to address questions regarding the long-term fate of these galaxies.
They found the outer ring such as that of the Cartwheel is formed 
$\sim$100-200~Myr after the collision. The ring would continue to expand and
eventually fade to form Giant Low Surface Brightness (GLSBs) galaxies 
after 0.5--1.0~Gyr of age. These simulations also predict the formation of a stellar bar triggered by
the galaxy interaction which can be long-lived (age$>$1~Gyr) and present in
the GLSB phase. {\it Malin 1}, a well known barred GLSB galaxy, is an
example of such a case \citep{2007AJ....133.1085B}.
The collision that formed the Cartwheel ring happened $\sim$300~Myr ago
\citep{2019IAUS..346..297W}, and hence the observed bar would not have had time to form after the collision.
Thus the observed strong bar in the Cartwheel was existing at the time of 
collision. Hence, the pre-collisional Cartwheel had a morphology of a barred 
late-type spiral galaxy with a small pseudobulge and a large disk. 

The Cartwheel clearly shows a 
strong negative colour gradient \citep{1992ApJ...399...57M}. Though a 
negative colour gradient is expected from an ageing population in the 
inner parts, the observed colour gradient can be quantitatively explained only if there is 
substantial contribution from the pre-collisional disc of old stars 
\citep{2001ApJ...554..281K}. The old stellar population was reported by 
\citet{1977MNRAS.178..473F} from spectroscopic observations by detecting the 
late-type stellar absorption lines in the observed spectrum. The study of colour 
modelling by \citet{2001A&A...377..835V} indicate that the pre-collision 
Cartwheel was a late-type spiral in which the old stellar components were embedded in 
an extended gaseous disc.

The old pre-collisional stellar components of the Cartwheel are likely made up of 
a  bar and an exponential bulge in the centre and a stellar disc extended 
over the entire size of the Cartwheel. 
Our finding suggests that the Cartwheel hosts a small exponential bulge which 
can be classified as pseudobulge given that the Sersic index $n<2$ 
\citep{2008AJ....136..773F}. 
\citet{2008ApJ...675.1194B} show that a significant fraction of barred late-type 
spiral galaxies host pseudobulges. Thus pseudobulges are common in late-type 
spiral galaxies. These type of bulges are believed to arise from disc material 
via secular evolution \citep{2004ARA&A..42..603K} induced either by bars or 
spiral structure \citep{1993A&A...271..391C, 2005MNRAS.358.1477A}. It has been 
observed that typically pseudobulges have younger stellar populations. 
However, it is not abnormal to have old pseudobulges in late-type spiral 
galaxies \citep{2007ApJ...658..960C}, for example, if by some mechanism the 
star formation in a galaxy is quenched after the formation of a pseudobulge, 
such bulge would quickly resemble inactive and old.

Many simulation studies show that the disc instabilities can lead to forming 
bars over a large range of disc masses that are stable for a long time. This also 
implies that bars are robust stellar structures, hence once formed, it is hard to 
destroy them \citep{1981A&A....96..164C, 1990ApJ...363..391P, 2004ApJ...604..614S, 
2005MNRAS.364L..18B, 2006ApJ...645..209D}. The fraction of gas present in disc 
plays a major role in the formation and evolution of bars as shown by 
\citet{2013MNRAS.429.1949A}. Strong bars are difficult to form in gas-rich disc 
galaxies compared to gas-poor ones. The fact that redder and gas-poor spiral 
galaxies have significantly increased bar fraction \citep{2012MNRAS.424.2180M} 
points towards a scenario in which the secular evolution of disc galaxies 
is driven by the growth of bars thus making the bar an important component for the 
evolution of galaxies. 

A stellar bar helping the formation of a pseudobulge during secular evolution
is another possibility. As shown in Table~\ref{tab:bartab} the bar in Cartwheel is massive compared to 
pseudobulge. This emphasises the fact that if pseudobulges are formed through 
disc instabilities through bars then on average pseudobulges are significantly 
less massive compared to bar for typical disc galaxies \citep{2011MNRAS.415.3308G}. 
Thus it is likely that the pre-collision Cartwheel was a barred late-type (later 
than SBc) spiral galaxy with a small pseudobulge, and possibly a large disc. 
The bar and the pseudobulge have survived the collision.

Cartwheel as an archetype collisional ring galaxy has been the subject of many 
simulation studies. In these simulations, it has been shown using some impact 
parameters how the morphology and star formation history of disc galaxy can be 
changed due to head-on collision by the intruder galaxy and produce ring structure 
\citep{1977ApJ...212..616T,  1990ApJ...358...99S, 1993MNRAS.261..804H, 
2003Ap&SS.284..499H, 2008AN....329..948M,  2012MNRAS.420.1158M, 2012MNRAS.425.2255F, 
2018MNRAS.473..585R}. The formation mechanisms of collisional ring galaxies are 
well constrained in these simulation studies given that the time scale to develop a
ring morphology and its lifespan is short compared to the Hubble time-scale. 
However, simulations that address the impact of such collisions on the survival
of central morphological components are rare.

Do bars survive in a collission that produces ring galaxies? 
This question was addressed by \citet{1997MNRAS.286..284A} in their 
N-body simulation study to create the rings in galactic disks by infalling 
small companions. They found that the bar survives the interaction, and slowly grows fatter 
developing into an oval structure. A small ring surrounds the bar, which 
expands and detaches the bar, forming an arm at one side of the bar (see their 
Figure~18). The observed HST-D morphology surrounding the bar resembles very 
much the results of their simulation. They further found that depending on the 
position of the impact, the bar shows an offset from the centre of mass of the 
galaxy (central bulge in our case). The photometric centre of our oval-shaped 
bar is shifted by 1.04\arcsec\  with respect to the bulge, a characteristic similar to that 
predicted in the N-body simulations. Using such N-body/SPH numerical simulations 
the dynamical effects of the interaction between an initially barred galaxy and 
a small companion was studied by \citet{2003MNRAS.341..343B}  and they found 
that the interactions can produce nuclear and circumnuclear disks (maybe a 
pseudobulge),  offset bars and tidal arms connected to the end of the bar. 
Based on the simulation results from \citet{2003MNRAS.341..343B} the fate of 
the bar in Cartwheel can be determined by the impact position and it can be 
hypothesised that in Cartwheel there may have been an impact on the bar such 
that the tidal force exerted on the bar during the impact does not disrupt the 
bar structure much and the stellar bar survived the impact. In other words, it 
is well known that the stellar bars are more robust structure and cannot be 
destroyed easily. Stellar bars can be weakened or destroyed in minor mergers, or 
by growing central mass concentration (CMC) or under the dynamical influence of 
Super Massive Blackholes (SMBHs) over long time scale (several gigayears) 
\citep{2003MNRAS.341..343B, 2004ApJ...604..614S, 2005MNRAS.363..496A, 
2013MNRAS.429.1949A}. This means that the head-on or off-axis collisions into 
the disc of the target galaxy by an intruder or companion galaxy that caused the 
star-formating ring are insufficient to destroy any pre-existing stellar bar.

Our discovery of the bar in the Cartwheel, makes Cartwheel the first 
collisionally formed ring galaxy where a bar has been detected. 
\citet{2006MNRAS.370.1607W} reported that the partial or C-shaped star-forming 
ring morphology observed in NGC\,922 which has an off-centred star-forming bar is 
a result of the slightly off-axis drop-through collision of a companion dwarf 
galaxy. This collision has enhanced the star formation in the ring and the 
central part of NGC\,922. \citet{2010AJ....139.1369P} reported many young star 
clusters of ages of less than 7~Myr and H$_\alpha$ emission in the partial or 
C-shaped star-forming ring. However, the morphology of NCG\,922 suggests it to be far from a classical
collisional ring galaxy.

The question of whether the bar was pre-existing or formed after the collision 
can be addressed directly through spatially resolved study of stellar 
populations in the bar region, using data from Integral Field Spectroscopy (IFS) 
such as MUSE. \citet{2019IAUS..346..297W} find super solar metallicity for the central region,
suggesting that the central structures were existing before the collision.
The discovery of bar in the Cartwheel also paves way to carry out 
near infrared (NIR) imaging observations of other collisional ring galaxies.

Large populations of collisional ring galaxies are expected at high redshift due to higher
rates of collisions and mergers. However, not many of them are discovered 
yet \citep[see e.g.][]{2006ApJ...651..676E}. Recently, \citet{2020arXiv200511880Y} has 
reported a discovery of a collisional ring galaxy at a redshift of $z=2.19$, which 
presents a new insight on ring formation and the evolution of disc galaxies in 
the early Universe. They reported a possible large bar (similar to the size of 
Milky Way's bar) embedded in a giant disc. Studies such as this would give
insight on the survival of bars under violent gas accretion events 
\citep{2012ApJ...758..136S,2012ApJ...757...60K}. 
{\it James Webb Space Telescope} (JWST) in the mid-IR wavelengths would be ideal
to carry out such studies \citep{2008MNRAS.389.1275D}.

To summarise this work, the discovery of bar and pseudobulge in Cartwheel 
puts impetus to include bars in future numerical and theoretical studies of 
collisional ring galaxies. The bar in Cartwheel is a strong and robust structure 
that survived the collision and remained unaffected by the impact. 
Thus, collisional ring galaxies provide opportunities for the study of a new, 
less explored aspect of bar evolution both observationally and theoretically.

\section*{Acknowledgements}

We thank Ivanio Puerari for his comments on an earlier version of this manuscript. We thank the anonymous referee whose insightful comments have improved both the content and presentation of this paper. This research has made use of the NASA/IPAC Extragalactic Database (NED), which is operated by the Jet Propulsion Laboratory, California Institute of Technology (Caltech) under contract with NASA. We acknowledge the usage of the HyperLeda database (http://leda.univ-lyon1.fr). This research has made use of the services of the ESO Science Archive Facility and based on observations collected at the European Southern Observatory under ESO programme 66.B-0666(B). 





\bibliographystyle{mnras}
\bibliography{ref} 








\bsp	
\label{lastpage}
\end{document}